\documentclass[preprintenumbers,prd,nofootinbib]{revtex4-2}%
\usepackage{amsmath}
\usepackage{amsfonts}
\usepackage{amssymb}
\usepackage{graphicx}
\usepackage{braket}
\begin{document}
\title{Inflationary evolution in quantum cosmology from FRLW supersymmetric models}
\author{N.E. Mart\'{\i}nez-P\'erez}
\email{nephtalieliceo@hotmail.com}
\author{C. Ram\'{\i}rez}
\email{cramirez@fcfm.buap.mx}
\affiliation{Benem\'erita Universidad Aut\'onoma de Puebla, Facultad de Ciencias
F\'{\i}sico Matem\'aticas, P.O. Box 165, 72000 Puebla, M\'exico.}
\author{V. V\'azquez-B\'aez}
\email{manuel.vazquez@correo.buap.mx}
\affiliation{Benem\'erita Universidad Aut\'onoma de Puebla, Facultad de Ingenier\'{\i}a, 72000 Puebla, M\'exico.}

\begin{abstract}
We consider inflationary scenarios for FRLW supersymetric models with a scalar field, whose Wheeler-DeWitt equation has analytic solutions. Following previous work of the authors, we analyse the setting of the scalar field as clock, leading to an effective time dependent wave function, and a time dependent mean value of the scale factor. We study several superpotentials, which lead to an evolution that corresponds to consistent inflationary scenarios. 
\end{abstract}

\maketitle

\section{Introduction}
\label{intro}
It is widely accepted, that the observable universe originates from a homogeneous early phase, beginning presumably around the Planck scale, just after a less understood phase of quantum spacetime. This homogeneous phase has a classical description by the FRLW model. On the other side, the quantization of inhomogeneous perturbations in a classical background leads to quantum fluctuations, whose evolution during inflation explains the seeds of structure of the universe \citep{mukhanov}. Thus, a quantum treatment seems to be natural for the homogeneous phase. 
Quantum cosmology gives a canonical quantization of general relativity  \citep{Hartle}. The hamiltonian constraint, generator of time reparametrizations, is implemented as a time independent Schr\"odinger equation, the Wheeler-DeWitt equation. Additionally, the solution of the WdW equation, the so called wave function of the universe, has to be supplemented by a way to obtain probabilities, considering the singular character of the universe. Further, as the hamiltonian operator acting on the wave function vanishes, this is a timeless theory. However, time must be reinstated in some way for the corresponding classical theory, with one of its components playing this role \citep{tiempo,Kuchar,Isham,Anderson}.

In recent years, there have been several proposals in standard quantum cosmology, which mainly relate time to the scale factor, essentially by a gauge fixing. These proposals involve mostly approximations of the type: semiclassical, Born-Openheimer with a weakly coupled matter sector, and a large scale factor. In \citep{barvisnky}, a general gauge fixing was analyzed, which has been applied in \citep{barvisnkyk} to several minisuperspace models, with a classical `extrinsic' time. In \citep{venturi98}, the Born-Oppenheimer approximation was used to separate the scale factor from a weakly coupled matter sector. It allows to define a time as in WKB approach, proportional to the square of the scale factor, and parametrizes matter evolution with an effective hamiltonian, see also \citep{moniz1}. Further, in \citep{venturi98} and a subsequent series of works, increasingly complexer settings have been analyzed, an inflaton with mass term in \citep{venturi98} , additional generic matter in \citep{venturi03}, Mukhanov-Sasaki scalar perturbations for a de Sitter evolution in \citep{venturi13}, tensor perturbations in a general slow-roll inflationary setting in \citep{venturi14}. The parameters of the last work have been restricted in \citep{venturi15} by comparison with observational data, and in \citep{venturi16} were considered  effects on the spectra of primordial perturbations. In \citep{venturi18} the consequences of the interference of the wave functions before and after a bounce are analyzed. For supersymmetric quantum cosmology, in \citep{moniz1,obregon2}, in the semiclassical approach, the effect of supersymmetry was explored on the properties of the solution. 

In \citep{previo:2016}, we have worked out a supersymmetric model that has analytic solution of the WdW equation. In the present work, with a slightly simpler WdW equation by a different operator ordering, we consider the time choice by an analysis of the wave function. The general expression of this wave function permits the identification of a curve of most probable values in configuration space (superspace), parametrized by the scalar field, and suggests to choose it as time. Such a choice has been known for a long time \citep{banks}. Here we analyse in detail the wave function regarding the choice made in \citep{previo:2016}, of an effective time dependent wave function, with a probability density that corresponds to the conditional probability of measuring a value of the scale factor, for a given value of the scalar field. With this wave function, we compute a time dependent mean value of the scale factor, which gives a trajectory that follows closely the previously mentioned curve of most probable values. Further, we consider the quantum cosmology of this approach, and explore two inflationary scenarios.

We consider supersymmetric models because, even if at LHC energies it has not been found to the present day, it might be broken at higher energies, and it continues to be part of important candidates of ultraviolet completions for the quantum theory of gravity, supergravity and string theory. Supersymmetry relates in a nontrivial way fermions and bosons, and in a supersymmetric quantum field theory the fermionic and bosonic divergences cancel \citep{wess}. 
Thus, supersymmetric quantum cosmology \citep{obregon0,eath,moniz} is a relevant option for the study of quantum cosmology. Supersymmetry can be formulated by the extension of spacetime translations to translations in a Grassmann, extended spacetime, which includes fermionic coordinates, called superspace\footnote{Should not be confused with the superspace of geometrodynamics.}. The fields on this supersymmetry-superspace are called superfields and supergravity can be formulated as a general relativity theory on a supermanifold \citep{wess,cupa}. There are several formulations for supersymmetric extensions of homogeneous cosmological models \citep{eath,moniz}. One class of such formulations comes from dimensional reduction of four or higher dimensional supergravity theories, by considering only time dependent fields, and integrating the space coordinates \citep{ryan}. The other class is obtained by supersymmetric extensions of homogeneous models, invariant under general reparametrizations of time \citep{graham,graham2}, or invariant under general reparametrizations on a superspace, with anticommutative coordinates besides time \citep{tkachwf,previo}.

Regarding quantization of the fermionic degrees of freedom in these theories, the fermionic momenta, as usual, are eliminated by solving the second class constraints, and the Dirac brackets are quantized in several ways. One way is representing the fermions by matrices \citep{obregon0,obregon1}, and another way is representing half of the fermion degrees of freedom by differential operators \citep{eath2}, or by creation operators acting on a vacuum state \citep{graham2,previo:2016}. Thus, the wave function is spinorial in the first case, or a finite expansion of possible states in the second case. In any case, the Wheeler-DeWitt equation is equivalent to a system of first degree differential equations. In simple cases these equations have exact solutions \citep{tkachwf,obregon1,previo}, and the integration constants can be assimilated in the normalization, i.e. no initial conditions are required, although the state will depend on the model. For supersymmetric extensions of higher order theories, like $f(R)$ theories, the differential equations might not be first order \citep{nephtali}.

In sections \ref{FRLW} and \ref{componentform}, we review the models, given in superfield formalism. In section \ref{quantization} we review the quantization of the models from \citep{previo:2016}, the supersymmetric Wheeler-DeWitt equations have an analytic solution, which depends on the scale factor and the superpotential. In section \ref{time} we make a discussion of the problem of time. The identification of high probability paths in configuration space, to which mean trajectories correspond, leads to the identification of the scalar field as time. Thus, following \citep{previo:2016}, a time dependent, effective wave function, can be given. This effective wave function allows to compute mean values of the scale factor, which give a time evolution. This scale factor is inversely proportional to the cubic root of the superpotential, and we obtain inflationary behavior from suitable exponential superpotentials, as shown in section \ref{inflation}. For an appropriate exit of inflation, we introduce a small constant, of the order $e^{-3{\cal N}}$, for ${\cal N}$ $e$-folds. We illustrate the inflationary behavior for two types of decreasing superpotentials. We discuss several cases, and plot the scale factor, the acceleration, and the comoving Hubble radius. Finally, in section \ref{discussion}, we make a short discussion with some remarks and future work perspectives. In an appendix we give the effective wave function and the scale factor for $k=1$.

%%%%%%%%%%%%%%%%%%%%%%%%%%%%%%%%%%%%%%%%%%%%%%%%%%%
%%%%%%%%%%%%%%%%%%%%%%%%%%%%%%%%%%%%%%%%%%%%%%%%%%%
\section{Supersymmetric FRLW model with a scalar field}
\label{FRLW}

The large scale observable universe has been studied in general relativity by the FRLW metric with scalar fields. This is a quite general setting that could follow from a fundamental theory, and can account for inflation, primordial matter generation and structure formation, and dark energy. We consider the most studied model, the simplest one, with one minimally coupled scalar field $\frac{1}{2\kappa^2}\int\sqrt{-g}Rd^4x+\int\sqrt{-g}[\frac{1}{2}\partial^\mu\phi\partial_\mu-V(\phi)]d^4x$. For the FRLW metric it reduces to the well known form, where $\kappa^2=\frac{8\pi G}{c^4}$,
\begin{equation}
I=\frac{1}{\kappa^{2}}\int\left\{  -\frac{3}{c^2}N^{-1}a\dot{a}^{2}+3Nka-Na^3\Lambda+\kappa^{2}a^{3}\left[\frac{1}{2c^2}N^{-1}\dot{\phi}^{2}-NV\left(\phi\right)\right]\right\}dt.
\label{actioncosmo}%
\end{equation}
This Lagrangian is invariant under general time reparametrizations. From this action follow the Friedmann equations and the conservation equation for a perfect fluid described by the scalar field $\phi(t)$, i.e. in natural units and comoving gauge
$\frac{\dot{a}^{2}}{a^{2}}-\frac{\Lambda}{3}+\frac{k}{a^2}=\frac{\kappa^{2}}{3}\rho$, $\frac{2\ddot{a}}{a}+\frac{\dot{a}^{2}}{a^{2}}-\Lambda+\frac{k}{a^2}=-\kappa^{2}p$ and
$\dot{\rho}+\frac{3\dot{a}}{a}(p+\rho)=0$,
with $\rho=\frac{1}{2} \dot{\phi}^{2}+V(\phi)$ the energy
density, and $p=\frac{1}{2} \dot{\phi}^{2}-V(\phi) $ the pressure for the
perfect fluid $\phi(t)$.
The momenta are $\pi_a=-\frac{6}{c^2\kappa^{2}}N^{-1}a\dot{a}$ and $\pi_\phi=-\frac{1}{c^2}N^{-1}a^3\dot{\phi}$
The Hamiltonian is $H=NH_0$, where $H_0$ is the hamiltonian  constraint, 
which generates time reparametrizations. 
%%%%%%%%%%%%%%%%%%%%%%%%%%%%%%%%
\subsection{Supersymmetric cosmology}
Supersymmetric cosmology can be obtained from one dimensional supergravity \cite{tkachwf}. Here we shortly review the derivation of the supersymmetric Wheeler-DeWitt equation following \citep{previo,previo:2016}. In these works, we have formulated it as general relativity on supersymmetry-superspace,
$t\rightarrow z^{M}=(t,\Theta,\bar{\Theta})$, where $\Theta$ and $\bar{\Theta}$
are anticommuting coordinates, the so called ``new'' $\Theta$-variables \citep{wess}. Hence, under $z^{M}\rightarrow z^{\prime M}=z^{M}+\zeta^{M}(z)$, the superfields,  see e.g. \cite{wess,cupa},
transform as $\delta_{\zeta}\Phi(z)=-\zeta^{M}(z)\partial_{M}\Phi(z)$,
and their covariant derivatives are $\nabla_{A}\Phi={\nabla_{A}^{\ M}(z)\partial_{M}}\Phi$. $\nabla_{A}^{\ M}(z)$ is the superspace vielbein,
whose superdeterminant gives the invariant superdensity
$\mathcal{E}=\mathrm{Sdet}{\nabla_{M}}^{A}$, $\delta_{\zeta}\mathcal{E}=(-1)^{m}\partial_{M}(\zeta^{M}\mathcal{E})$. 
For the supersymmetric extension of the FRLW metric, in the covariant Wess-Zumino gauge \cite{cupa}, we have $\mathcal{E}=-N-\frac{i}{2}(\Theta\bar{\psi}+\bar{\Theta}\psi)$ \cite{previo}. In this formulation, to the scale factor and the scalar field correspond real scalar superfields \citep{wess,tkachwf}. 
\begin{align}
\mathcal{A}\left(  t,\Theta,\bar{\Theta}\right)  =a\left(  t\right)
+\Theta{\lambda}\left(  t\right) -\bar{\Theta}\bar\lambda\left(  t\right)
+  \Theta\bar{\Theta}B\left(t\right),\\
\Phi\left(  t,\Theta,\bar{\Theta}\right)  =\phi\left(  t\right)  +\Theta{\eta}\left(  t\right)
-\bar{\Theta}\bar\eta(t)+\Theta\bar{\Theta}G\left(  t\right). \label{superfields}%
\end{align}
Here scalar has the usual meaning, of invariant under time reparametrizations. Note that under time reparametrizations, all the components of these superfields, i.e. $a(t)$, $\lambda(t)$, $\bar{\lambda}(t)$, $B(t)$, and $\phi(t)$, $\eta(t)$, $\bar{\eta}(t)$, $G(t)$, are scalar.

The supersymmetric extension of the action (\ref{actioncosmo}), for $k=0,1$, is
$I=I_G+I_{M}$ , where $I_{G}$ is the supergravity action, and $I_M$ is the matter term \cite{tkachwf,superfield,previo:2016}
\begin{align}
I_{G}&=\frac{3}{\kappa^{2}}\int\mathcal{E}\left(  \mathcal{A}\nabla
_{\bar{\Theta}}\mathcal{A}\nabla_{\Theta}\mathcal{A}-\sqrt{k}\mathcal{A}^{2}\right)  d\Theta d\bar{\Theta}dt, \label{susyraction}\\
I_{M}&=\int\mathcal{E}\mathcal{A}^{3}\left[  -\frac
{1}{2}\nabla_{\bar{\Theta}}\Phi\nabla_{\Theta}\Phi+W\left(  \Phi\right)
\right]  d\Theta d\bar{\Theta}dt, \label{susymaction}%
\end{align}
where $W$ is the superpotential.
%%%%%%%%%%%%%%%%%%%%%%%%%%%%%%%%%%%%%%%%%%%%%%%%%%%%%%%%%%%%%%%%%%%%%%%%%
%%%%%%%%%%%%%%%%%%%%%%%%%%%%%%%%%%%%%%%%%%%%%%%%%%%%%%%%%%%%%%%%%%%%%%%%%
\section{Component formulation}
\label{componentform}
The component action follow from (\ref{susyraction}) and (\ref{susymaction}),  after performing the Grassmann integrals, and solving the equations of motion of the auxiliary fields $B$ and $G$, see \citep{previo:2016}. Then, after the redefinitions
$\lambda\rightarrow a^{1/2}\lambda$, $\bar{\lambda}\rightarrow a^{1/2}\bar{\lambda}$,
$\eta\rightarrow a^{3/2}\eta$, and $\bar{\eta}\rightarrow a^{3/2}\bar{\eta}$
\begin{align*}
L_{Tot} &  =-\frac{3a\dot{a}^{2}}{c^2N\kappa^{2}}+\frac{a^{3}\dot{\phi}^{2}}{2c^2N}
+\frac{3kNa}{\kappa^{2}}+3\sqrt{k}Na^{2} W+\frac{3N\kappa^{2}}{4}a^{3}W^{2}-\frac{1}{2}a^{3}N{W'}^{2}
+\frac{3i}{c\kappa^{2}}\left(  \lambda\dot{\bar{\lambda}}+\bar{\lambda}\dot{\lambda}\right)
-\frac{i}{2c}\left(  \eta\dot{\bar{\eta}}+\dot{\eta}\bar{\eta}\right) \\
& +\frac{3\sqrt{a}\dot{a}}{c\kappa^2N}\left(  \psi\lambda-\bar{\psi}\bar{\lambda}\right)
-\frac{a^{3/2}\dot{\phi}}{2cN}\left(  \psi\eta-\bar{\psi}\bar{\eta}\right)
+\frac{3i\dot{\phi}}{2c}\left(  \lambda\bar{\eta}+\bar{\lambda}\eta\right)
+3N\left(\frac{\sqrt{k}}{\kappa^{2}a}-\frac{3}{2}W\right)\lambda\bar{\lambda}
+N\left(-\frac{3\sqrt{k}}{2a}+\frac{3\kappa^{2}}{4}W-W''\right)\eta\bar{\eta}  \\
& +3ia^{3/2}\left(\frac{\sqrt{k}}{\kappa^2a}-\frac{1}{2}W\right)\left(  \psi\lambda+\bar{\psi}\bar{\lambda}\right)
-\frac{ia^{3/2}}{2}W'\left(  \psi\eta+\bar{\psi}\bar{\eta}\right)
-\frac{3N W'}{2}\left(  \lambda\bar{\eta}-\bar{\lambda}\eta\right)
-\frac{3}{2N\kappa^2}\psi\bar{\psi}\lambda\bar{\lambda}+\frac{1}{4N}\psi\bar{\psi}\eta\bar{\eta},
\end{align*}
where $W\equiv W(\phi)$,  $W'\equiv \partial_\phi W(\phi)$, and $W''\equiv \partial_\phi^2 W(\phi)$.
The Hamiltonian is $H=NH_{0}+\frac{1}{2}\psi S-\frac{1}{2}\bar{\psi}\bar{S}$, where $H_0$ is the hamiltonian constraint and $S$, and $\bar{S}$ are the supersymmetric constraints \citep{previo:2016}. The canonical  Dirac Brackets are $\{a,\pi_{a}\}=\{\phi,\pi_{\phi}\}=1$,
$\{\lambda,\bar{\lambda}\}_{+}=\frac{c\kappa^2}{6}$, $\{\eta,\bar{\eta}\}_{+}=-c$, where $\{,\}_{+}$ are fermionic Dirac brackets. 
The constraints satisfy 
\begin{align}
\{S,\bar S\}_{+}&=-2H_0,\label{ss}\\
\{H_0,S\}&=\{H_0,\bar S\}=0. \label{hs}
\end{align}
Regarding time reparametrizations, the three constraints $H_0$, $S$, and $\bar{S}$ are scalar.

The scalar potential in the hamiltonian $H_0$ is \citep{previo:2016}
\begin{equation}
V_S=\frac{3\sqrt{k}}{a}W-\frac{3\kappa^{2}}{4}W^{2} +\frac{1}{2}{W'}^{2}.\label{scalarpot}
\end{equation}
Note that for $k=0$, the sign of the superpotential does not matter for the scalar potential.
%%%%%%%%%%%%%%%%%%%%%%%%%%%%%%%%%%%%%%%%%%%%%%%%%%%%%%%%%%%
%%%%%%%%%%%%%%%%%%%%%%%%%%%%%%%%%%%%%%%%%%%%%%%%%%%%%%%%%%%
\section{Quantization}
\label{quantization}
Homogeneous cosmology is a mechanical system, hence it can be quantized by the methods and with the interpretation of conventional quantum mechanics. However, there are several well known problems. The first one is that this is a system that cannot be observed from the outside, hence it is not possible to perform repeated measurements in identical, observer shaped, conditions \citep{halliwell}. Nevertheless,  observables like the scale factor, can be measured by a set of observations. Other problem is that the hamiltonian vanishes, and a time variable cannot be introduced by means of the Schr\"odinger equation. However,  the Wheeler-DeWitt equation gives a time independent Schr\"odinger equation with zero eigenvalue, whose solution can be interpreted as the probability amplitude for the universe, and which depends on the metric and matter degrees of freedom. Quantum mechanics in the Schr\"odinger picture tells us that the observables are represented by time independent operators, whose eigenvalues are the allowed values of these observables. The fact that the theory does not give a time evolution is a well known consequence of invariance under time reparametrizations. It has been argued that time is an internal property that can be determined by a the choice of a clock \citep{Kuchar}. On the other side, the observed universe is classical \citep{halliwell}, hence its description is given by mean values of the quantum operators. We further discuss the time problem in section \ref{time}.
%%%%%%%%%%%%%%%%%%%%%%%%%%%%%%%%%
\subsection{Supersymmetric Wheeler-DeWitt equations}
For the derivation of the supersymmetric Wheeler-DeWitt equations, we follow \citep{previo:2016}, but with somewhat different conventions. We choose a slightly different operator ordering for fermions, which gives simpler solutions. For consistency it is required that the Hamiltonian operator is hermitian, hence the supercharges must satisfy $\bar S=S^\dagger$ and  $S=\bar{S}^\dagger$.The non-zero (anti)commutators are
\begin{equation}
[a,\pi_{a}]=[\phi,\pi_{\phi}]=i\hbar,\qquad\{\lambda,\bar{\lambda}\}=\frac{4\pi}{3}l_P^2,\qquad\{\eta,\bar{\eta}\}=-\hbar c,  \label{conmutadores}
\end{equation}
where $l_P^2=\frac{\hbar G}{c^3}$ is the Planck length.  
For the quantization, we redefine the fermionic degrees of freedom as
$\lambda=\sqrt{\frac{\hbar c\kappa^2}{6}}\alpha$, $\bar\lambda=\sqrt{\frac{\hbar c\kappa^2}{6}}\bar\alpha$, $\eta=\sqrt{\hbar c}\beta$
and $\bar\eta=\sqrt{\hbar c}\bar\beta$. Hence the anticommutators are
\begin{equation}
\qquad\{\alpha,\bar{\alpha}\}=1,\qquad\{\beta,\bar{\beta}\}=-1.  \label{conmutadoresf}
\end{equation}
as well as $\alpha^{2}=\beta^{2}=\bar{\alpha}^{2}=\bar{\beta}^{2}=0$.
The bosonic momenta are represented by derivatives, $\alpha$ and $\beta$ are 
annihilation operators, and $\bar{\alpha}$ and $\bar{\beta}$ are creation operators.
We fix the ordering ambiguities by Weyl ordering, which for fermions is antisymmetric. Hence
\begin{align}
\frac{1}{\sqrt{\hbar c}} S  & =\frac{c\kappa}{2\sqrt{6}}\left(a^{-\frac{1}{2}}\pi_a+\pi_a a^{-\frac{1}{2}}\right)\alpha
+ca^{-\frac{3}{2}}\pi_\phi\beta
+\frac{3i\kappa}{\sqrt{6}}  a^\frac{3}{2}W\alpha
+i a^\frac{3}{2}W'\beta
-i\frac{\sqrt{6k}}{\kappa}a^\frac{1}{2}\alpha
-\frac{i\sqrt{3}}{4\sqrt{2}}\hbar c\kappa a^{-\frac{3}{2}}\alpha[\bar{\beta},\beta],\label{Squant}\\
\frac{1}{\sqrt{\hbar c}} \bar S  & =\frac{c\kappa}{2\sqrt{6}}\left(a^{-\frac{1}{2}}\pi_a+\pi_a a^{-\frac{1}{2}}\right)\bar\alpha
+ca^{-\frac{3}{2}}\pi_\phi\bar\beta
-\frac{3i\kappa}{\sqrt{6}}  a^\frac{3}{2}W\bar \alpha
-i a^\frac{3}{2}W'\bar \beta
+i\frac{\sqrt{6k}}{\kappa}a^\frac{1}{2}\bar \alpha
+\frac{i\sqrt{3}}{4\sqrt{2}}\hbar c\kappa a^{-\frac{3}{2}}\bar\alpha[\bar{\beta},\beta].\label{Sbquant}
\end{align}
The anticommutator  $\{S,\bar S\}=-2\hbar c H_0$ gives the quantum hamiltonian
\begin{align}
H_0=&-\frac{c^2\kappa^2}{24}\left(a^{-1}\pi_a^2+\pi_a^2a^{-1}\right)+\frac{c^2}{2}a^{-3}\pi_\phi^2
-\frac{\sqrt{3} i}{2\sqrt{2}}\hbar c^2\kappa a^{-3}\pi_\phi(\alpha\bar\beta+\bar\alpha\beta)-\frac{3k}{\kappa^2}a
-\frac{\sqrt{k}}{4}\hbar ca^{-1}[\alpha,\bar\alpha]+\frac{3\sqrt{k}}{4}\hbar ca^{-1}[\beta,\bar\eta]\nonumber\\
&-\frac{3\kappa^2}{4}a^3W^2+3\sqrt{k}a^2W+\frac{1}{2}a^3{W'}^2+\frac{3}{8}\hbar c\kappa^2W[\alpha,\bar\alpha]
-\frac{3}{8}\hbar c\kappa^2W[\beta,\bar\beta]+\frac{\sqrt{3}}{2\sqrt{2}}\hbar c\kappa W'(\alpha\bar\beta-\bar\alpha\beta)
+\frac{1}{2}\hbar cW''[\beta,\bar\beta]\nonumber\\
&+\frac{3}{16}(\hbar c\kappa)^2a^{-3}\left(\bar\alpha\alpha\beta\bar\beta+\alpha\bar\alpha\bar\beta\beta\right).\label{Hquant}
\end{align}
The Hilbert space is generated from the vacuum state $\ket{1}$, which
satisfies $\alpha\ket{1}=\beta\ket{1}=0$. Hence, there are four orthogonal states
\begin{equation}
\ket{1},\quad\ket{2}=\bar{\alpha}\ket{1},\quad\ket{3}=\bar{\beta}\ket{1}\quad
\mathrm{and}\quad\ket{4}=\bar{\alpha}\bar{\beta}\ket{1},\label{estados}
\end{equation}
which have norms $\braket{2|2}=\braket{1|1}$, $\braket{3|3}=-\braket{1|1}$ and
$\braket{4|4}=-\braket{1|1}$. Hence, a general state will have the form
\begin{equation}
\ket{\psi}=\psi_{1}(a,\phi)\ket{1}+\psi_{2}(a,\phi)\ket{2}+\psi_{3}(a,\phi)\ket{3}+\psi_{4}(a,\phi)\ket{4}.\label{estado}
\end{equation}
Therefore, from the constraint equation $S\ket{\psi}=0$, we get
\begin{align}
a\left(\partial_{a}-\frac{3}{\hbar c} a^{2}W+\frac{6\sqrt{k}}{\hbar c\kappa^2}a+\frac{1}{2}a^{-1}\right)  \psi_{2}
-\frac{\sqrt{6}}{\kappa}\left(  \partial_{\phi}-a^{3}W'\right)  \psi_{3}=0,\label{ec23}\\
\left(  \partial_{a}-\frac{3}{\hbar c} a^{2}W+\frac{6\sqrt{k}}{\hbar c\kappa^{2}}a-a^{-1}\right)  \psi_{4}=0\quad \text{and}\quad
\left(  \partial_{\phi}-\frac{1}{\hbar c}a^{3}W'\right)  \psi_{4}=0,\label{ec4}
\end{align}
while from $\bar{S}\psi=0$
\begin{align}
 a\left(\partial_{a}+\frac{3}{\hbar c} a^{2}W-\frac{6\sqrt{k}}{\hbar c\kappa^2}a+\frac{1}{2}a^{-1}\right)  \psi_{3}
-\frac{\sqrt{6}}{\kappa}\left(  \partial_{\phi}+a^{3}W'\right)  \psi_{2}=0,\label{ec32}\\
\left(  \partial_{a}+\frac{3}{\hbar c} a^{2}W-\frac{6\sqrt{k}}{\hbar c\kappa^{2}}a-a^{-1}\right)  \psi_{1}=0,\quad \text{and}\quad
\left(  \partial_{\phi}+\frac{1}{\hbar c}a^{3}W'\right)  \psi_{1}=0,\label{ec1}
\end{align}
Note that the terms with $a^{-1}$ in (\ref{ec23})-(\ref{ec1}) differ from those in  in \citep{previo:2016} due to a different operator ordering in the Hamiltonian. In fact, the classical hamiltonian in \citep{previo:2016} has a term $a^{-1}\pi_a^2$ that can be ordered in many ways to give a hermitian operator as $a^{-1}\pi_a^2\rightarrow \frac{1}{2k+l}(ka^{-1}\pi_a^2+l\pi_a ^{-1} \pi_a+k\pi_aa^{-1})$. 
%%%%%%%%%%%%%%%%%%%%%%%%%%%%%%%
\subsection{Solutions}\label{sec_sol}
As the Wheeler-DeWitt equation is second order, its solutions require boundary conditions. However, in the supersymmetric theory the equations are first order (\ref{ec23})-(\ref{ec1}), and have unique solutions, which can be fixed by consistency and normalization.
The equations for $\psi_{1}$ and $\psi_{4}$ can be straightforwardly solved yielding the, up to constant factors, unique solutions \citep{tkachwf}
\begin{align}
\psi_{1}(a,\phi) &  =a\exp \left[-\frac{1}{\hbar c} \left( a^{3}W(\phi)-\frac{3\sqrt{k}a^{2}}{\kappa^{2}} \right) \right]  ,\label{psi1aT}\\
\psi_{4}(a,\phi) &  =a\exp \left[\frac{1}{\hbar c} \left(  a^{3}W(\phi)-{\frac{3\sqrt{k}a^{2}}{\kappa^{2}}}\right)\right]  .\label{psi4aT}%
\end{align}
For the equations (\ref{ec23}) and (\ref{ec32}), the solutions for $W(\phi)=0$ and $k=0$ are
$\psi_2(a,\phi)=e^{\frac{1}{2a}}[f_{+}(ae^{\kappa\phi/\sqrt{6}})+f_{-}(ae^{-\kappa\phi/\sqrt{6}})]$ and  $\psi_3(a,\phi)=e^{\frac{1}{2a}}[f_{+}(ae^{\kappa\phi/\sqrt{6}})-f_{-}(ae^{-\kappa\phi/\sqrt{6}})]$, where $f_\pm$ are arbitrary functions. These solutions are not defined at $a=0$, unless they are trivial.
Thus  \citep{previo:2016}, we choose the solutions
$\ket{\psi}=C_1\,\psi_{1}(a,\phi)\ket{1}+C_4\,\psi_{4}(a,\phi)\ket{4}$,
where the factors are arbitrary constants.
The norm of this state is
\begin{equation}
\braket{\psi|\psi}=\left[|C_1|^2\int|\psi_1(a,\phi)|^2dad\phi-|C_4|^2\int|\psi_4(a,\phi)|^2dad\phi\right]\braket{1|1}.\label{norma}
\end{equation}
Classically $a\geq0$, and could be a problem for quantization, see e.g. \cite{Isham}. It could require an infinite wall. However, the solutions (\ref{psi1aT}) and (\ref{psi4aT}) already vanish at $a=0$. For a positive superpotential, $\psi_{1}$ has a bell form and tends to zero as $a$ increases, see figure \ref{fig_campana}. In this case the solution $\psi_{2}$ must be set the trivial one. Oppositely, for a negative superpotential, $\psi_{2}$ tends to zero as $a$ increases, and $\psi_{1}$ must be discarded. 
For $\phi\to\pm\infty$, the behavior of  (\ref{psi1aT}) and (\ref{psi4aT}) depends on the form of the superpotential. 
Therefore, we consider only positive or negative superpotentials, and we choose $\braket{1|1}=1$ for positive superpotentials, and $\braket{1|1}=-1$ for negative superpotentials. Hence
\begin{align}
\ket{\psi}  &  = C\psi_1(a,\phi)\ket{1},\quad\text{if}\ W(\phi)>0,\label{psi1w}\\
\ket{\psi}  &  =C\psi_4(a,\phi)\ket{4},\quad\text{if}\ W(\phi)<0. \label{psi4w}
\end{align}
From the expansion (\ref{estado}), and (\ref{estados}), we see that these states correspond to scalars. By construction, these states are invariant under supersymmetry transformations. 
Usually, a wave function corresponds to a localized particle, with well defined position probabilities, and probability conservation. These conditions also guarantee hermiticity of operators. On the other side, free particles cannot be localized in a finite volume, and their wave functions do not vanish at infinity, but can be compared.
If we restrict the superpotential to be an even function of $\phi$, then any of the two states (\ref{psi1w}) or (\ref{psi4w}) is an even function of $\phi$, and the operators $\pi_\phi$ and $H_0$ are self-adjoint, even if the wave function does not vanish at $\phi\to\pm\infty$. In the following we will consider only such superpotentials. The self-adjoint Hamiltonian constraint is consistent with the lack of evolution, in the Heisenberg picture
\begin{equation}
\bra{\psi}\frac{da}{dt}\ket{\psi}=\frac{i}{\hbar}\bra{\psi}[H_0,a]\ket{\psi}=0.\label{noevol}
\end{equation}
In the following, unless otherwise stated, we will consider $k=0$. In this case we can write (\ref{psi1w}) and (\ref{psi4w}) as
\begin{align}
\psi(a,\phi)  = Ca\exp \left[-\frac{1}{\hbar c} a^{3}|W(\phi)| \right].\label{psiw}\\
\end{align}
This wave function differs slightly from the one in \citep{previo:2016}, by the power of the $a$ in front of the exponential, due to a different operator ordering, as mentioned in the preceding section.

In the appendix we give the expressions for $k=1$.

%%%%%%%%%%%%%%%%%%%%%%%%%%%%%%%%%%%%%%%%%%%%%%%%%%%%%%%%%%%%%%%%%%%%%
%%%%%%%%%%%%%%%%%%%%%%%%%%%%%%%%%%%%%%%%%%%%%%%%%%%%%%%%%%%%%%%%%%%%%
\section{Time}
\label{time}
Ordinary quantum mechanics assigns operators with real spectra to observables, and probability amplitudes for their occurrence. A time dependence of the wave function is generated by the Hamiltonian operator. However, in a real occurring state, time requires that there is a certain energy indeterminacy. Hence, the energy spectrum must have at least two values, and time dependent mean values arise by interference among different energy states. Thus, if the hamiltonian is a constraint and vanishes, time is undetermined. Otherwise, if there are different energy states, there must be transitions among them, hence there is an environment. If we mean by universe everything, there is no environment. However, we could consider a scalar component of the universe as time; this component must be present in the theory \citep{banks}, see also \citep{Kuchar,Anderson}. 

We have taken in \citep{previo:2016} the inflaton as time, and we have proposed a time dependent effective wave function, from which time dependent mean values are computed. Here we analyse the wave function (\ref{psiw}) and its probability density regarding the motivation of this choice of time. 

We consider the conventional interpretation of quantum mechanics for the solutions of the Wheeler-DeWitt equation. Hence the square module of the wave function must give probabilities for all possible three-geometries. Further, invariance under time reparametrizations ensures that the superspaces (in the sense of geometrodynamics) of each of the space slices are equivalent. Thus, this wave function describes, in quantum mechanical sense, the space geometries of the whole space-time. 
Furthermore, if it is possible to identify mean trajectories in superspace along regions around maxima of the wave function, then it should be possible to parametrize these trajectories following the idea of Misner's supertime \citep{misner}. These trajectories should be around classical trajectories \citep{halliwell}, corresponding to effective theories. Strictly speaking, measurements should give random values around these mean values. 

\begin{figure}[h]
\centering
\includegraphics[height=4cm,width=6cm]{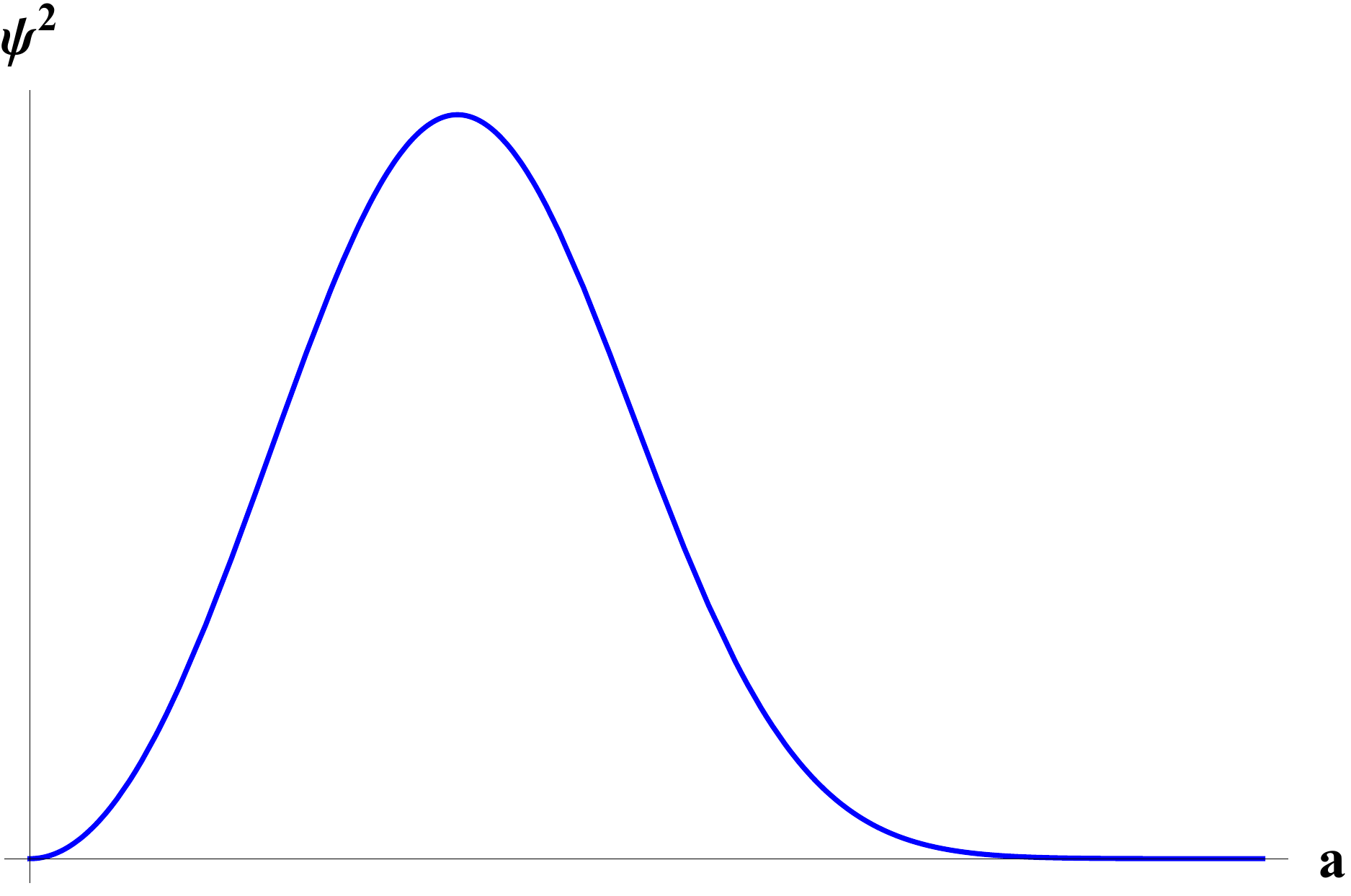}
\caption{{\protect\footnotesize {Profile of $\psi^2(a,\phi)$, for $\phi$ constant.}}}
\label{fig_campana}
\end{figure}

In the model of this paper, the effective configuration space is given by the scale factor and the scalar field. Further, from  (\ref{psiw}) we see, that if we keep $\phi$ constant, the probability density of the wave function (\ref{psiw}) has the generic bell form of  figure \ref{fig_campana}, with the maximum at
\begin{equation}
a_{\text{max}}(\phi)=\left[\frac{\hbar c}{3W(\phi)}\right]^{1/3}.\label{amax}
\end{equation}
Hence, variating $\phi$, these maxima will form in the plane $(a,\phi)$, a curve of most probable values of the scale factor. The probability density along this curve is
\begin{equation}
\psi^2(a_{\text{max}},\phi)=e^{-2/3} a^2_{\text{max}}(\phi),
\end{equation}
Therefore, higher values of the scale factor along this curve will have a higher relative probability, and we could think of an expanding universe for this wave function. On the other side, from the quantum theory we obtain in general time dependent mean values, and we can speculate that later times correspond to higher probabilities. Hence, time would increase monotonically with the scale factor. Nevertheless, $a_{\text{max}}$ is driven by the scalar field (\ref{amax}), hence a natural choice for time is the scalar field, and the superpotential should ensure that $a_{\text{max}}$ increases properly. The choice made corresponds to a gauge, where the scalar field is constant on the spacial slices. Actually, as it is scalar, the values of the scale factor are not affected by a change of time gauge.
Otherwise, under the previous assumptions, the universe will not be localized in the $\phi$ direction. Hence, the wave function is not normalizable in this direction, but we can consider relative probabilities in this direction, i.e. quotients of probabilities. As the value of $\phi$ is highly uncertain, a measurement gives random values, and we can ask then for the probability that the scale factor takes a value, which is given by the conditional probability of obtaining this value of $a$, given a value of $\phi$
\begin{equation}
\left\vert\Psi(a,\phi)\right\vert^2=\frac{\left\vert\psi(a,\phi)\right\vert^2}{\int_0^\infty da\,\vert\psi(a,\phi)\vert^2}.\label{condprob}
\end{equation}
This probability must be used if there is a correlation between both fields, as required for a clock in \citep{banks}, where it is called relative probability.
The probability (\ref{condprob}) can be obtained  \citep{previo:2016} from a wave function normalized for a given value of $\phi$
\begin{equation}
\Psi(a,\phi)=\frac{1}{\sqrt{\int_0^\infty da\,\vert\psi(a,\phi)\vert^2}}\,\psi(a,\phi),
\label{onda}
\end{equation}
i.e. $\int_0^\infty da\,\vert\Psi(a,\phi)\vert^2=1$.
Thus, making $\phi\to t$, the probability density is
\begin{equation}
\left\vert\Psi(a,t)\right\vert^2=\frac{\left\vert\psi(a,t)\right\vert^2}{\int_0^\infty da\,\vert\psi(a,t)\vert^2}.
\end{equation}
With the solution (\ref{psiw}), (\ref{onda}) becomes
\begin{equation}
\Psi(a,t)=\sqrt{\frac{6|W(t)|}{\hbar c}}a\exp\left[{-\frac{a^{3}|W(t)|}{\hbar c}}\right],
\label{ondaw}
\end{equation}
and satisfies the conservation equation $\frac{\partial}{\partial t}|\Psi(a,t)|^2
+\frac{1}{3}\frac{\partial}{\partial a}\left[a\frac{d\ln|W(t)|}{dt}|\Psi(a,t)|^2\right]=0$ \citep{previo:2016}.
Further, as the observed universe is classical, what we can give a meaning to, following Ehrenfest theorem, is mean values. Thus, under the preceding ansatz, the time dependent setup arises from mean values with the probability amplitude (\ref{onda}). For the scale factor we get \citep{previo:2016}
\begin{equation}
a(t)=\int_0^\infty a|\Psi(a,t)|^2 da=\Gamma\left(4/3\right)\left[\frac{\hbar c}{2|W(t)|}\right]^{1/3}, \label{amean}
\end{equation}
which is close to (\ref{amax}), as their quotient is $\Gamma(4/3)(3/2)^{1/3}\sim 1.02$.

The quantum fluctuations produce standard deviations
\begin{equation}
(\Delta a)^2=\left[\Gamma\left(5/3\right)-\Gamma\left(4/3\right)^2 \right]\left[\frac{\hbar c}{2|W|}\right]^{2/3}, \label{sigmaa}%
\end{equation}
and
\begin{equation}
(\Delta\pi_a)^2=\hbar^2\, \Gamma\left(1/3\right)\left[\frac{2|W|}{\hbar c} \right]^{2/3},\label{SigmaPia}
\end{equation}
from which follows the uncertainty relation \citep{previo:2016}
\begin{equation}
\Delta a\Delta\pi_a=\sqrt{\Gamma\left(1/3\right)\left[\Gamma\left(5/3\right) -\Gamma\left(4/3\right)^{2}\right]}\, \hbar
\approx0.53\,\hbar.\label{incertidumbre}
\end{equation}
We are not making any approximation. As the fermionic degrees of freedom do not have classical counterparts, these mean values do not necessarily correspond to trajectories that approximate classical solutions. See e.g. \citep{obregon2}. Actually, (\ref{amean}) can be inserted into the Friedmann equations, from which a potential can be read out
\begin{equation}
\frac{1}{3c^4\kappa^2}\left(\frac{2{W'}^{2}}{W^2}-\frac{W''}{W}\right),
\end{equation}
which differs from the scalar potential (\ref{scalarpot}).

The results for this section can be given also analytically for $k=1$, see Appendix. They involve Hypergeometric, AiryBi, and AiryBi$'$ functions, which have exponential behaviour, and their numerical evaluation is troublesome. As we are here interested on qualitative features, we will restrict ourselves to $k=0$. In this case, considering that the sign of the superpotential does not have consequences for the wave function, nor for the scalar potential, we will choose the superpotential as positive definite in the following.
%%%%%%%%%%%%%%%%%%%%%%%%%%%%%%%%%%%%%%%%%%%%%%%%%%%%%%%%%%%%%%%%%%%%%
%%%%%%%%%%%%%%%%%%%%%%%%%%%%%%%%%%%%%%%%%%%%%%%%%%%%%%%%%%%%%%%%%%%%%
\section{Inflationary scenarios}
\label{inflation}
In this section we will consider several superpotentials that lead to cosmological scenarios. The scalar field has real values,  from $-\infty$ to $\infty$, and as mentioned in section \ref{sec_sol}, we require that the superpotential is an even function and always positive. To describe the origin of the universe, it is convenient to choose time beginning at $t=0$. Hence, these superpotentials correspond to bouncing universes.
Thus, we consider universes with a defined origin at t=0, described by homogeneous quantum cosmology from its very beginning, although above the Planck scale, we would require full quantum gravity. 

We will consider superpotentials depending on the initial value of the scalar potential. First $a(0)=0$, hence $\lim_{t\to0}W(t)=\infty$ (\ref{amean}), and the wave function (\ref{onda}) will vanish at $t=0$, $\Psi(a,0)=0$, similar to figure \ref{fig_campana}. Hence, the universe will arise for $t>0$ from nothing. Otherwise, for finite $W(0)$, we have $a(0)=\Gamma\left(4/3\right)\left[\frac{\hbar c}{2|W(0)|}\right]^{1/3}$. We can obtain superpotentials of the first sort by $W(t)\to t^{-2\alpha}U(t)$, with $\alpha$ positive, and $U(0)$ finite.

From (\ref{amean}) we get
\begin{equation}
\ddot{a}(t)=\frac{a(t)}{9}\left[4\frac{{\dot{W}}^2(t)}{W^2(t)}-3\frac{\ddot{W}(t)}{W(t)}\right].
\end{equation}
Hence the acceleration is positive if
\begin{equation}
{\dot{W}}^2(t)>\frac{3}{4}W(t)\ddot{W}(t) \label{winf}
\end{equation}
We consider inflation starting when the scale factor acceleration becomes positive, at $t=t_i$, until the exit when it becomes zero again $\ddot{a}(t_e)=0$, with ${\cal N}=\ln\frac{a(t_e)}{a(t_i)}=\frac{1}{3}\ln\frac{W(t_i)}{W(t_e)}$ $e$-folds. Note that in this way, as the scale factor is scalar, ${\cal N}$ is independent of the time parametrization.

An indicative of the feasibility of inflation, is the comoving Hubble length
\begin{equation}
(aH)^{-1}=-\frac{3\times2^{1/3}}{\Gamma(4/3) (c\hbar)^{1/3}}\frac{W(t)^{4/3}}{\dot{W}(t)}.\label{horizon}
\end{equation}
In order to have an effective model in more familiar terms, we can substitute the scale factor in the Friedmann equations to obtain the energy density and the pressure for $k=0,1$, as well as a time dependent potential, $V(t)=\rho(t)-p(t)/c^2$; this potential does not coincide with the scalar potential (\ref{scalarpot}).

For simplicity, in the following we will consider an adimensional time $t\to\tau=t/t_p$, in Planck time units, $t_p=\sqrt{\hbar G/c^5}$ is the Planck time. However, the dot will continue to denote time derivatives, $\dot{f}\equiv\frac{df}{dt}$.

%%%%%%%%%%%%%%%%%%%%%%%%%%%%%%%%%%%%%%%%%%%%%%%%%%%%%%%%%%%%%%%%%%%%%%%%%%%%
\subsection{Superpotentials}
We will consider superpotentials that can lead to an inflationary universe, which satisfy (\ref{winf}). One such superpotential has been considered in \citep{obregon2}, an exponential superpotential $W(\phi)\sim e^{-C\phi}$. We have explored this superpotential, and it seems that it does not produce enough $e$-folds. Thus, we will restrict the analysis to two types of superpotentials, which show the main desirable features:

Gaussian superpotentials
\begin{equation}
W(\tau)=\frac{c^4 M_p^3}{\hbar^2}\tau^{-2\alpha}\left( e^{-\tau^2}+\lambda\right),
\label{gaussian}
\end{equation}

and Step superpotentials
\begin{equation}
W(\tau)=\frac{c^4 M_p^3}{\hbar^2}\tau^{-2\alpha}\left(\frac{1}{\tau^{2}+\lambda}+1\right),\label{step}
\end{equation}
where $M_p=\sqrt{\frac{\hbar c}{G}}$ is the Planck mass, $\alpha$ positive integer, and $\lambda$ positive constant.
We consider representative cases which reproduce an inflationary period with ${\cal N}\sim 60$ $e$-folds, evaluated considering the beginning and exit of inflation from the time interval where acceleration is positive, and a consistent comoving Hubble length. As we will see, a very small $\lambda$ is required for enough $e$-folds, and we set $\lambda=e^{-\nu}$. Regarding other parameters, as the tensor-scalar ratio, an estimate considering an effective FRLW cosmology with one scalar, leads to inconsistent results, a more precise evaluation is needed.

Initially, for these superpotentials, the behaviour is as follows:  for $\alpha=0$ the scale factor has a value $a(0)>0$ with a very small positive acceleration, and  for $\alpha\geq 1$, $a(0)=0$ with a vanishing acceleration.
In order to see if there is a consistent inflation, we evaluate the values of the scale factor at $\tau=0$, and the corresponding acceleration. If $a(0)>0$,
then inflation begins at the time $\tau=\tau_i$, when the acceleration becomes positive, until it becomes negative, at $\tau=\tau_f$.
If the acceleration is positive already at $\tau=0$, inflation begins right away.
Further, if $a(0)=0$ with positive acceleration from the beginning, then at the exit there are infinite $e$-folds, unless we discard an initial full quantum period. Otherwise, if $a(0)=0$ and $\ddot{a}(0)<0$, then inflation will begin as soon as the acceleration becomes positive, at $\tau_i$, and finish when the acceleration becomes negative again, at $\tau_f$, with ${\cal N}=\ln\frac{a(\tau_f)}{a(\tau_i)}$. In the following we will consider natural units $c=1$ and $\hbar=1$, i.e. $M_P=1/\sqrt{G}$, and the Planck and time length $l_P=t_P=M_P^{-1}$.

%%%%%%%%%%%%%%%%%%%%%%%%%%%%%%%%%%%%%%%%%%%%%%%%%%%%%%%%%%%%%%%%%%%%%%%%%%%%%%%
\subsection{Gaussian superpotentials}
\begin{figure}[h]
\centering
\includegraphics[width=8.6cm]{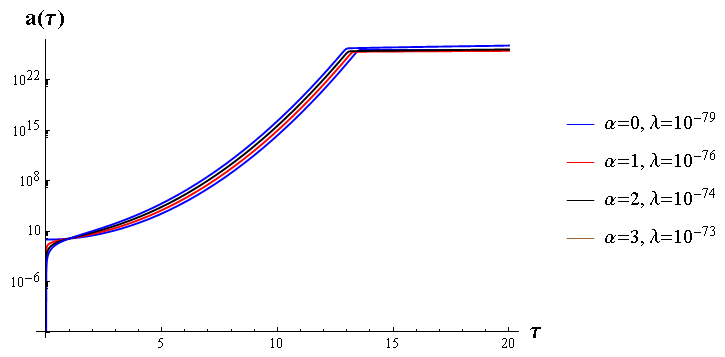}
\caption{{\protect\footnotesize {Scale factor $a(\tau)$ for gaussian superpotentials, the graphics are logarithmic for clarity.}}}
\label{fig_ga}
\end{figure}

\begin{figure}[h]
\centering
\includegraphics[width=8.6cm]{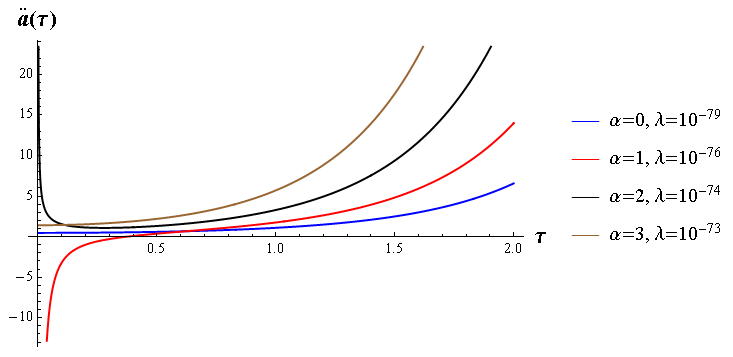}
\caption{{\protect\footnotesize {Acceleration $\ddot{a}(\tau)$ for early times, gaussian superpotentials.}}}
\label{fig_gac1}
\end{figure}

\begin{figure}[h]
\centering
\includegraphics[width=8.6cm]{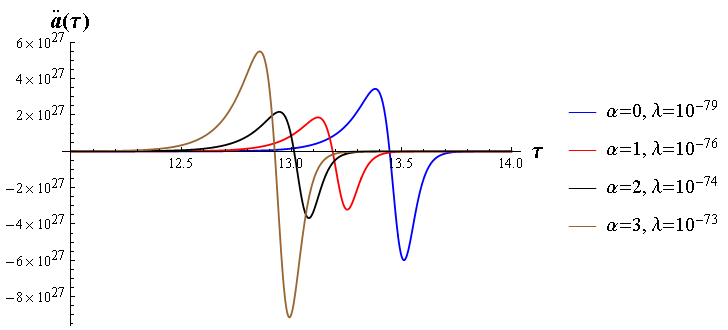}
\caption{{\protect\footnotesize {Acceleration $\ddot{a}(\tau)$ for gaussian superpotentials.}}}
\label{fig_gac}
\end{figure}

\begin{figure}[h]
\centering
\includegraphics[width=8.6cm]{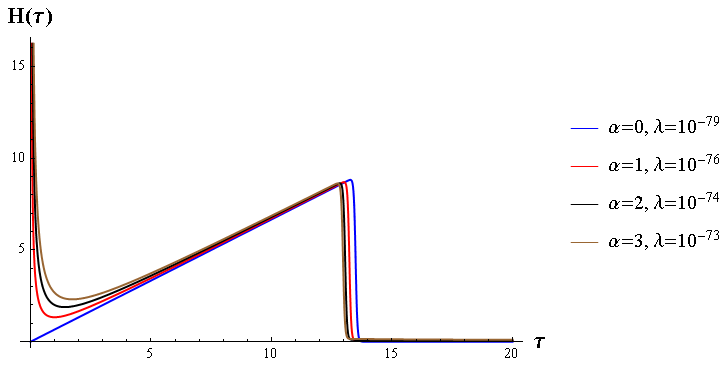}
\caption{{\protect\footnotesize {Hubble parameter $H(\tau)$ for gaussian superpotentials.}}}
\label{fig_gH}
\end{figure}

\begin{figure}[h]
\centering
\includegraphics[width=8.6cm]{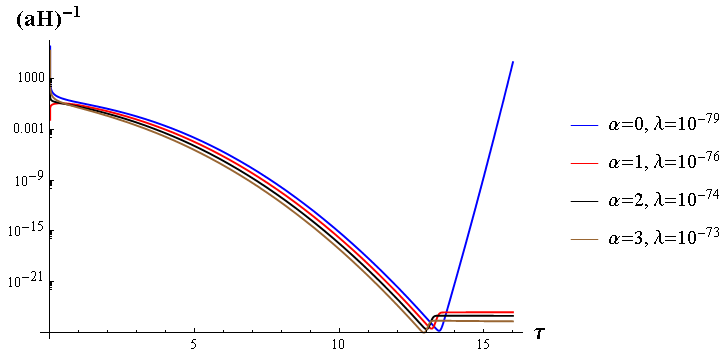}
\caption{{\protect\footnotesize {Comoving Hubble radius $(aH)^{-1}$ for gaussian superpotentials, the graphics are logarithmic for clarity.}}}
\label{fig_gh}
\end{figure}

The scale factor for (\ref{gaussian}) is
\begin{equation}
a(\tau)=\frac{\Gamma(4/3)\tau^{2\alpha/3}l_P}{2^{1/3} \left(e^{-\tau^2}+\lambda \right)^{1/3}}.\label{agauss0}
\end{equation}
Therefore, for $\alpha=0$, neglecting $\lambda$ with respect to 1, we have, figure \ref{fig_ga}
\begin{equation}
a_0=a(0)=\frac{\Gamma(4/3)l_P}{2^{1/3}(1+\lambda)^{1/3}}\approx\frac{\Gamma(4/3)l_P}{2^{1/3}},\label{a00}
\end{equation}
with a positive acceleration, figure \ref{fig_gac1}
\begin{equation}
\ddot{a}(0)\approx\frac{2^{2/3}\Gamma(4/3)M_P}{3}.\label{gacc}
\end{equation}
On the other side, for later times, the acceleration vanishes around the value at which the superpotential (\ref{gaussian}) becomes nearly constant, at  $e^{-\tau^2}\approx\lambda$, i.e. $\tau\approx\sqrt{\nu}$, see figure \ref{fig_gac}, and the scale factor becomes 
\begin{equation}
a_f\approx\frac{\Gamma(4/3)l_P}{2^{1/3}\lambda^{1/3}},\label{af}
\end{equation}
Thus, for $\alpha=0$, the $e$-fold number satisfies
\begin{equation}
{\cal N}\lesssim\ln\frac{a_f}{a_0}=-\frac{1}{3}\ln\lambda=\frac{\nu}{3}.\label{efolds}
\end{equation}
Hence, a value $\lambda\sim10^{-79}$ is necessary for $\sim 60$ $e$-folds.
From (\ref{a00}), we obtain that $a(0)\sim 0.7\, l_P$, with an initial velocity $\dot{a}(0)=0$. The initial acceleration is $\ddot{a}(0)\sim 0.5\, M_P$, hence inflation begins right away, until the acceleration slows and becomes negative at
$\tau_f\approx 13.4$, see figure \ref{fig_gac}. We get $N=\ln\frac{a(\tau_f)}{a_0}\approx60.1$, consistently with (\ref{efolds}). The Hubble parameter is plotted in figure \ref{fig_gH}, and the comoving Hubble length in figure \ref{fig_gh}.

For $\alpha>0$, $a(0)=0$, and at early times, $\tau\ll 1$, the scale factor can be approximated by
$a(\tau)\approx\frac{\Gamma(4/3)l_P}{2^{1/3}}\tau^{2\alpha/3}e^{\frac{\tau^2}{3}}$,  and
$\ddot{a}(\tau)\approx \frac{1}{9}\Gamma(4/3)2^{2/3} \alpha (2\alpha-3) \tau^{2\alpha/3 - 2}M_P $. For $\alpha=0,1,2,3$ the acceleration s shown, for early times in \ref{fig_gac1}, and for late times in figure \ref{fig_gac}. Thus we have:

For $\alpha=1$, $\ddot{a}(0)=-\infty$, then grows and for $\lambda\sim 10^{-76}$, the acceleration becomes positive at
$\tau_i\sim 0.4$, with ${\cal N}=\ln a(\tau_f)/a(\tau_i)\sim 60.2$.

For $\alpha=2$, $\ddot{a}(0)=\infty$, decreases almost to zero for $0<\tau\lesssim 0.4$, and then grows again. Hence as $a(0)=0$, ${\cal N}$ is unbounded. However, the initial acceleration is very quickly decreasing, see figure \ref{fig_gac1}, and we could discard this initial period. In this case, if we consider inflation starting at $\tau\sim 0.4$, we get ${\cal N}\sim 60.2$.

For $\alpha=3$, $\ddot{a}(0)\simeq 1.4\, M_P$, and we have a situation similar to $\alpha=2$. If we consider inflation starting at $\tau\sim 0.5$, we get ${\cal N}\sim 60.5$.

In \ref{fig_ga}, we show the plots of the scale factor for the previous cases in logarithmic scale. Even if the previous results are alike, the comoving Hubble lengths, plotted in figure \ref{fig_gh}, are quite different.

\begin{figure}[h]
\centering
\includegraphics[width=8.6cm]{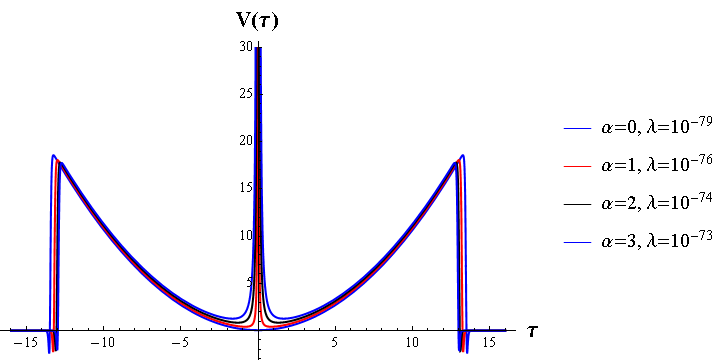}
\caption{{\protect\footnotesize {Effective potentials from Friedmann equations, for $k=0$, and $\alpha=0,1,2,3$ the potential becomes $\infty$ at $\tau=0$.}}}
\label{fig_potg0}
\end{figure}

With the previous results, we can compute the effective matter density and pressure for a perfect fluid. It turns out that the resulting potential differs from the scalar potential following from the hamiltonian (\ref{scalarpot}).  Moreover, for the gaussian potential and $k=0$,  in the analysed cases, the fluid has phantom energy. The effective potentials for $k=0$  are given in figure \ref{fig_potg0}.

As we can see from figures \ref{fig_gh} and \ref{fig_sh}, Hubble's radius decreases during a time interval consistent with that of inflation solving the planicity problem, Hubble's radius must decrease during inflation so that we have less structures causally connected for those times and, subsequently, during the big-bang evolution, Hubble's radius grows and more structures are causally reconnected again \citep{baumann}. We must note that different combinations for the values of $\alpha$ and $\lambda$ give similar behaviours but some correspond to a slightly bigger radius and thus to a longer inflationary period, allowing us to tune the duration of inflation according to data. The minimum, along with an adequate neighbourhood, in Hubble's radius for each curve could be interpreted as a reheating epoch once some matter is introduced in a more realistic model.
%%%%%%%%%%%%%%%%%%%%%%%%%%%%%%%%%%%%%%%%%%%%%%%%%%%%%%%%%%%%%%%%%%%%%%%%%%%
\subsection{Step superpotentials}
\begin{figure}[h]
\centering
\includegraphics[width=8.6cm]{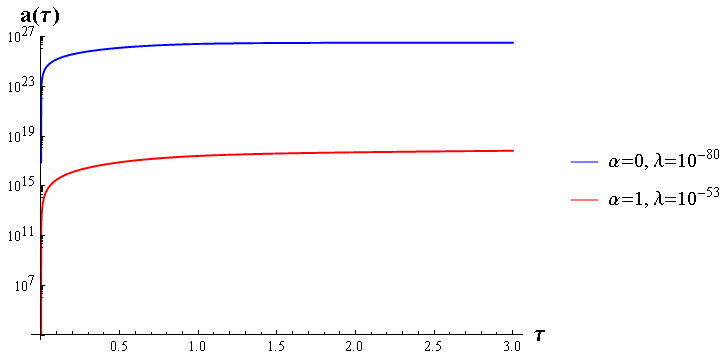}
\caption{{\protect\footnotesize {Scale factor $a(\tau)$ for step superpotentials.}}}
\label{fig_sa}
\end{figure}

\begin{figure}[h]
\centering
\includegraphics[width=13cm]{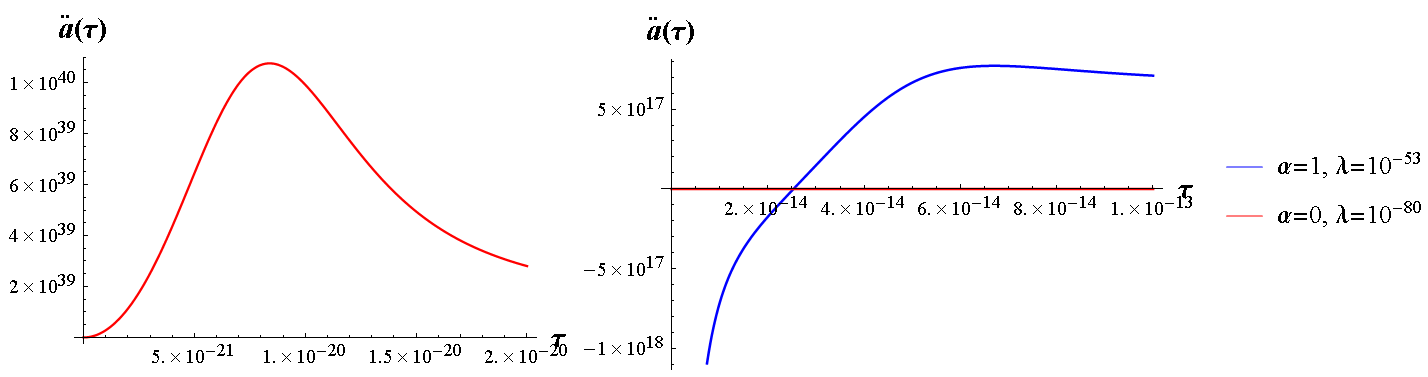}
\caption{{\protect\footnotesize {Initial acceleration for step superpotentials.}}}
\label{fig_sa1}
\end{figure}

\begin{figure}[h]
\centering
\includegraphics[width=13cm]{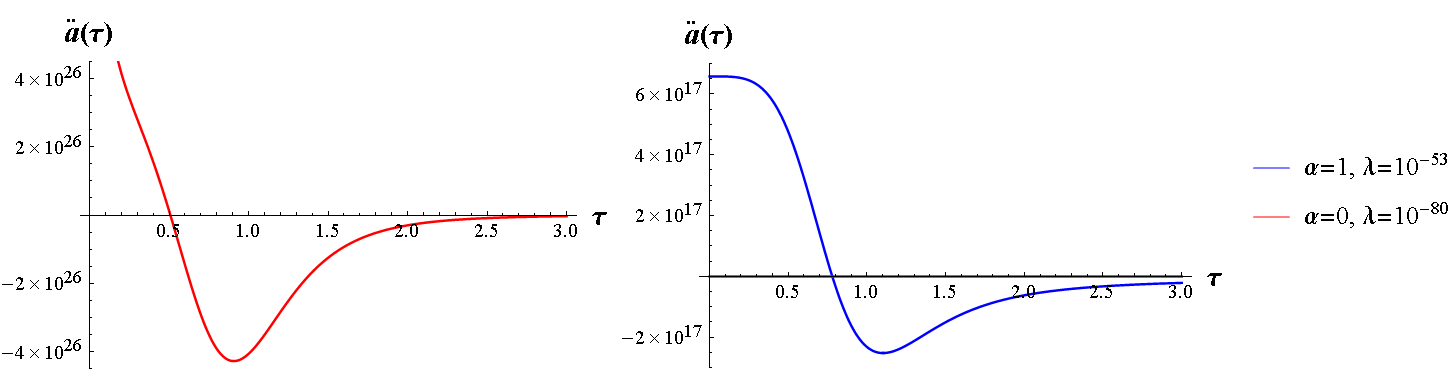}
\caption{{\protect\footnotesize {Acceleration $\ddot{a}(\tau)$ for step superpotentials.}}}
\label{fig_sac}
\end{figure}

\begin{figure}[h]
\centering
\includegraphics[width=13cm]{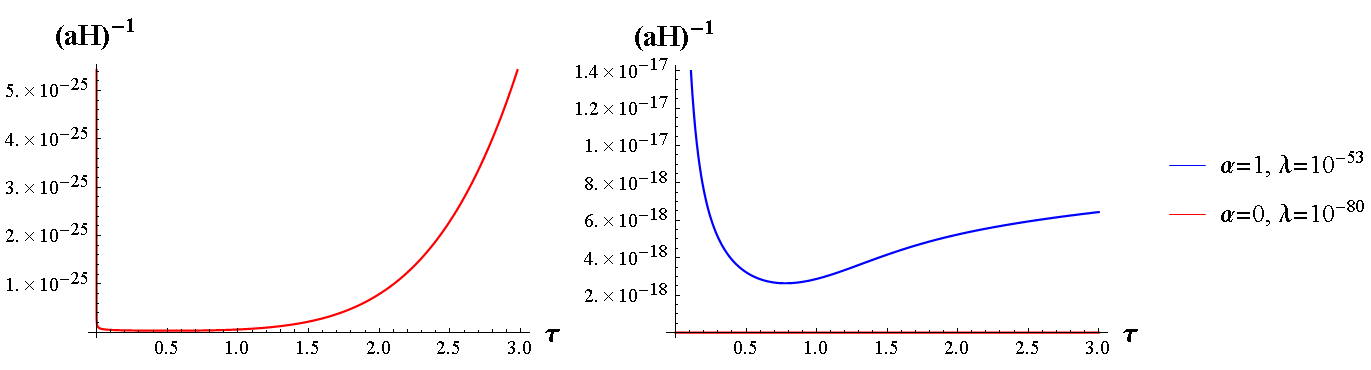}
\caption{{\protect\footnotesize {Comoving Hubble radius $(aH)^{-1}$ for step superpotentials.}}}
\label{fig_sh}
\end{figure}

\begin{figure}[h]
\centering
\includegraphics[width=8.6cm]{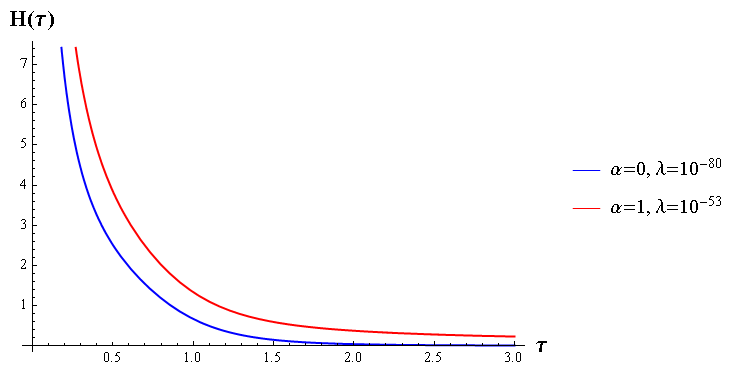}
\caption{{\protect\footnotesize {Hubble parameter $H(\tau)$ for step superpotentials.}}}
\label{fig_sH}
\end{figure}

For the step superpotentials (\ref{step}), we have
\begin{equation}
a(\tau)=\frac{\Gamma(4/3)l_P\tau^{2\alpha/3}}{2^{1/3} \left[(\tau^{2}+\lambda)^{-1}+1\right]^{1/3}}\label{astep}
\end{equation}
As well as for gaussian superpotentials, for $\alpha=0$ the scale factor initial value is (\ref{a00}), and for large time values, $\tau\gg 1$, the scale factor tends to (\ref{af}), figure \ref{fig_sa}.
Hence, the $e$-folds satisfy as well (\ref{efolds}).
The situation is similar as for the gaussian superpotential, the initial velocity and acceleration are zero, $\dot{a}(0)=0$, $\ddot{a}(0)=0$, and inflation begins from this moment, and ends at $\tau\approx 0.51$, when the acceleration becomes negative. For ${\cal N}\sim 60$ a value $\lambda\sim 10^{-80}$ is required. In figures \ref{fig_sa1}, \ref{fig_sac}, \ref{fig_sh}, and \ref{fig_sH}, the acceleration, comoving Hubble length, and Hubble parameter are plotted.

For $\alpha=1$ the the initial acceleration is minus infinite, and becomes positive at $\tau\sim 10^{-14}$, and further negative at $\tau\sim 3.8$, see figures \ref{fig_sa1} and \ref{fig_sac}. The 60 $e$-folds are reached for $\lambda\sim 10^{-53}$. The comoving Hubble length and the Hubble parameter are plotted in figures \ref{fig_sh} and \ref{fig_sH}.
Note that the time scales are different than for previous case. 

%%%%%%%%%%%%%%%%%%%%%%%%%%%%%%%%%%%%%%%%%%%%%%%%%%%%%%%%%%%%%%%%%%%%%%%%%%%%%%%%%%%%
%%%%%%%%%%%%%%%%%%%%%%%%%%%%%%%%%%%%%%%%%%%%%%%%%%%%%%%%%%%%%%%%%%%%%%%%%%%%%%%%%%%%
\section{Discussion}
\label{discussion}

Quantum cosmology gives a canonical quantization of general relativity in the Schr\"odinger picture, with the Wheeler-DeWitt equation as  time independent Schr\"odinger equation. The wave function gives the probability amplitudes for the occurrence of all possible spatial 3-geometries and field distributions, at any time. On the other side, the observed universe is classical and has time, and classical physics follows from quantum physics, hence an effective time evolution should follow from the quantum description. Further, the Wheeler-DeWitt equation is a second order scalar equation. In the supersymmetric case there is a system of first order equations, whose solutions are spinorial wave functions, see e.g. \citep{eath,moniz}. A particular meaning of the components of these wave functions has not been given; it would quickly lead to a Machian discussion. Frequently, there are only two nonvanishing components, see e.g. \citep{obregon1}, given by real exponentials with opposite exponent sign. 
Here we have considered a FRLW theory with a minimally coupled scalar field. It is well known, that single field models are extremely effective to account for the inflationary era. 
The solution of the corresponding supersymmetric Wheeler-DeWitt equation (\ref{ec23})-(\ref{ec1}) has four components, two of them have analytic expressions (\ref{psi1aT}),(\ref{psi4aT}), and only one tends to zero as $a\to\infty$. As the equations are homogeneous, and (\ref{psi1aT}),(\ref{psi4aT}) decouple, the other components can be taken to be trivial. The form of the solution (\ref{psiw}), suggests the ansatz that, in a certain gauge, time can be given by the scalar field, and trajectories for the observables are mean values on an effective wave function that gives conditional probabilities given a value of the scalar (\ref{amean}). Thus, for the scale factor we get an evolution $a(t)$, and we can perform time reparametrizations, considering that it is scalar $a'(t')=a(t)$. The resulting evolution does not correspond to a perturbative setting. These trajectories are classical if the quantum fluctuations are negligible.  It remains the question of how far can we get classical trajectories in this way for late times. Note that the time proposed here does not follow from a semiclassical approach, although it is not a quantum variable either. There is not a time operator.

We have considered in this formulation several inflationary scenarios, with superpotentials (\ref{gaussian}) and (\ref{step}). These superpotentials have a factor $\tau^{-2\alpha}$ which, for $\alpha\neq 0$, gives a wave function that vanishes for $\tau=0$, and produces a path of increasing probabilities, which could be seen as a time direction.  The inflation exit is implemented in the superpotentials through a constant $\lambda\sim e^{-3{\cal N}}$, where ${\cal N}$ are the required $e$-folds.
For each type of superpotential we considered several cases, plotted in the graphics \ref{fig_gac}-\ref{fig_sa1}. From these results we can expect that this sort of models could be explored for a realistic scenario, perhaps with two scalar fields, as in \citep{previo}. Moreover, inhomogeneous perturbations should be considered, to compute the evolution of the fluctuations. An interesting perspective presents the study of this formalism for dark energy.

\vskip 1truecm
\centerline{\bf Acknowledgments} C.R. thanks O. Obreg\'on and H. Garc\'{\i}a-Compe\'an, and  V. V.  thanks G. Garc\'{\i}a-Arroyo and J.A. V\'azquez,
 for interesting discussions.
 We thank VIEP-BUAP, CONACYT, and PRODEP-SEP for support. 

%%%%%%%%%%%%%%%%%%%%%%%%%%%%%%%%%%%%%%%%%%%%%%%%%%%%%%%%%%%%%%%%%%%%%%

\section{Appendix}
\label{appendix}
In this appendix we give,  for $k=1$, the expressions for the normalization factor for the wave function (\ref{psi1aT}) for $k=1$, and the time dependent scale factor (\ref{amean}).
The normalization factor of
\begin{equation}
\psi(a,\phi)=|C|a\exp\left[\frac{1}{\hbar c}\left(  -a^{3}|W(\phi)|+\frac{3\sqrt{k}a^{2}}{\kappa^{2}}\right)\right],\label{fonda1}
\end{equation}
is given by
\begin{align}
|C|^{-2}=&\frac{c\hbar}{6} \int_{-\infty}^{\infty}\frac{1}{|W(\phi)|}\Bigg[\, _2F_2\left(\frac{1}{2},1;\frac{1}{3},\frac{2}{3};\frac{8}{c\hbar\kappa^6 W^2(\phi)}\right)
+4\pi\left[\frac{2}{9c \hbar \kappa ^6W(\phi)^2}\right]^{1/3}e^{\frac{4}{c\hbar\kappa^6 W^{2}(\phi)}}\nonumber\\
&\times\bigg[\text{Bi}'\left(\left[\frac{6}{c \hbar \kappa ^6W(\phi)^2}\right]^{2/3}\right)
+\left[\frac{6}{c \hbar \kappa ^6W(\phi)^2}\right]^{1/3}\text{Bi}\left(\left[\frac{6}{c \hbar \kappa ^6W(\phi)^2}\right]^{2/3}\right)\bigg]\Bigg]d\phi, \label{C_1}
\end{align}
where $Bi$ is the Airy function of second kind. 

For the denominator of (\ref{onda})
\begin{align*}
\int_0^\infty da\,\vert\psi(a,\phi)\vert^2=&\frac{c \hbar }{18 W(\phi )} 
\times\Bigg\{3\text{  }\, _2F_2\left(\frac{1}{2},1;\frac{1}{3},\frac{2}{3};-\frac{8}{c\hbar \kappa ^6   W(\phi)^2}\right)
+4\times 6^{1/3} \pi  \left[\frac{1}{c \hbar \kappa ^6W(\phi )^2}\right]^{2/3}e^{-\frac{4}{c \hbar  \kappa ^6 W(\phi )^2}}\\
&\times\Bigg[6^{1/3} \text{Bi}\left(\left[\frac{6}{c\hbar \kappa ^6W(\phi )^2}\right]^{2/3}\right)
- \left[c \hbar \kappa ^6W(\phi )^2\right]^{1/3} \text{Bi}'\left(\left[\frac{6}{c \hbar \kappa ^6W(\phi)^2}\right]^{2/3}\right)\Bigg]\Bigg\},
\end{align*}
from which follows
\begin{align*}
a(t)=\Bigg\{9 \sqrt[3]{3} \kappa ^4 (c \hbar )^{2/3} W(t)^{4/3} \, _2F_2\left(1,\frac{3}{2};\frac{2}{3},\frac{4}{3};\frac{8}{c \kappa^6 \hbar  W(t)^2}\right)
e^{-\frac{4}{c\hbar  \kappa ^6  W(t)^2}}\\
+2^{2/3} \pi \text{  }\left[24+c \kappa ^6 \hbar  W(t)^2\right] \text{Bi}\left(\left[\frac{6}{c \hbar \kappa ^6W(t)^2}\right]^{2/3}\right)\\
+8 \sqrt[3]{2} 3^{2/3} \pi  \kappa ^2 \sqrt[3]{c \hbar } W(t)^{2/3}\text{  }\text{Bi}'\left(\left[\frac{6}{c \hbar \kappa ^6W(t)^2}\right]^{2/3}\right)\Bigg\}/\\
\Bigg\{3 \sqrt[3]{3} \kappa ^6 (c \hbar )^{2/3} W(t)^{7/3} \, _2F_2\left(\frac{1}{2},1;\frac{1}{3},\frac{2}{3};\frac{8}{c \kappa ^6 \hbar  W(t)^2}\right)
e^{-\frac{4}{c\kappa ^6 \hbar  W(t)^2}}\\
+4 \sqrt[3]{2} 3^{2/3} \pi  \kappa ^4 \sqrt[3]{c \hbar } W(t)^{5/3}\text{  }\text{Bi}'\left(\left[\frac{6}{c \hbar \kappa ^6W(t)^2}\right]^{2/3}\right)\\
+12\ 2^{2/3} \pi  \kappa ^2 W(t)\text{  }\text{Bi}\left(\left[\frac{6}{c \hbar \kappa ^6W(t)^2}\right]^{2/3}\right)\Bigg\}.
\end{align*}
Due to the exponential behaviour of the hypergeometric and Airy functions, which appear in the numerator and denominator in these expressions, it is difficult to handle them numerically.
\vskip 2truecm
%%%%%%%%%%%%%%%%%%%%%%%%%%%%%%%%%%%%%%%%%%%%%%%%%%%%%%%%%%%%%%%%%%%%%%%%%%

%%%%%%%%%%%%%%%%%%%%%%%%%%%%%%%%%%%%%%%%%%%%%%%%%%%%%%%%%%%%%%%%%%%%%%%%%%%%%%%%%%%%%%%%%%%%%%%
%%%%%%%%%%%%%%%%%%%%%%%%%%%%%%%%%%%%%%%%%%%%%%%%%%%%%%%%%%%%%%%%%%%%%%%%%%%%%%%%%%%%%%%%%%%%%%%

\end{document}